\documentclass[aps,preprint]{revtex4}
\usepackage{epsfig}
\usepackage{amsmath}
\usepackage{epsfig}

\begin{document}
\title
{\bf Position-momentum uncertainty products}    
\author{Zafar Ahmed$^{1}$ and Indresh Yadav$^2$}
\email{1:zahmed@barc.gov.in, 2:iykumarindresh288@gmail.com}
\affiliation{$^{1}$Nuclear Physics Division, Bhabha Atomic Research Centre, Mumbai 400085, India,\\
$^{2}$Solid State Physics Division, Bhabha Atomic Research Centre, Mumbai 400085, India}
\date{\today}
\begin{abstract} 
We point out two interesting features of position-momentum uncertainty product: $U=\Delta x \Delta p$.
We show that two special (non-differentiable)  eigenstates of the Schr{\"o}dinger operator with the Dirac Delta potential $[V(x)=-V_0 \delta(x)],V_0>0$, also satisfy the Heisenberg's uncertainty principle by yielding $U> \frac{\hbar}{2}$. One of these eigenstates
is a zero-energy and zero-curvature bound state.
\end{abstract}
\maketitle
That the position $(x)$ and momentum ($p$) of a particle, in quantum-world, cannot be measured precisely and simultaneously in the same direction is called Heisenberg's uncertainty principle. This is one of the most important features of the quantum-world and in the realm of the Schr{\"o}dinger equation it is  precisely stated  [1-3,6-10] as
\begin{equation}
\Delta x \Delta p \ge \frac{\hbar}{2},
\end{equation}
as the commutator $[x,p]=i\hbar$.
The uncertainty $\Delta A$ in an observable corresponding to an operator $A$ for an energy eigenstate $\psi(x)$ is defined as
\begin{equation}
\Delta A = \sqrt{<\psi|A^2|\psi> - <\psi|A|\psi>^2}.
\end{equation}
The equality sign in (1) is well known to occur for the ground state, $\psi_0(x)=A e^{-\alpha x^2}$, of the one dimensional harmonic oscillator. For the ground state, $\psi_0(x)=A \sin(\pi x/a)$, of the  well known infinitely deep well (IDW) potential  this product turns out to be $\frac{\hbar}{2} \sqrt{\frac{\pi^2-6}{3}}$ [3] which is  (approximately $0.5678\hbar$) a little more than $\hbar/2$ and it does not depend on the value of the width of the well. In textbooks usually these two potentials are discussed for the  uncertainty product. We shall be denoting the uncertainty product (1) as $U_{\psi(x)}$ as it is a property of the eigenstate.

In this Letter, we point two interesting features of $U$. Firstly, for the potentials which possess finite 
number of discrete bound states, $U$ admits the minimum value of $\frac{\hbar}{2}$ when their depth tends to infinity. Secondly, $U_{\psi(x)}=U_{\phi(x)}$, where $\phi(p)$ is the Fourier-transform of $\psi(x)$. Then we obtain $U$ for two eigenstates of Schr{\"o}dinger operator with Dirac delta potential. Both of these eigenstates are non-differentiable at $x=0$ and
and finding $U$ for them becomes tricky as acknowledged in Ref. [6]. More interestingly, one of them is a novel zero-energy and zero-curvature bound state [11,12]. One may wonder whether these eigenstates would satisfy the uncertainty principle by yielding $U>\frac{\hbar}{2}$.

Using special higher order functions, beautiful expressions of the uncertainty products for the  the exactly solvable symmetric Rosen-Morse potential, $V_{SRM}(x)=s(s+1) \tanh^2 (x)$ [2],
and the Morse oscillator, $V_M(x)=\lambda^2(1-e^{-x})^2$ [2],  have been obtained [4] in terms of poly-gamma function, $\Psi^\prime(z)$ [5] (not to be confused with the wave function, $\psi(x)$). For the ground state of $V_{SRM}$, we have [4]
\begin{equation}
\Delta x \Delta p=\frac{\hbar}{2} \sqrt{\frac{s^2 \Psi^\prime(s)}{s+1/2}}, \quad s>0.
\end{equation}
For the Morse oscillator we have [4]
\begin{equation}
\Delta x \Delta p= \frac{\hbar}{2} \sqrt{ (2\lambda -1) \Psi^\prime(2\lambda-1)}, \quad \lambda >1/2.
\end{equation}
For large values of $z$, $\Psi^\prime (z) \sim \frac{1}  {z}+\frac{1}{2z^2}$. Also we have a recurrence relation $\Psi^\prime(z+1)=\Psi^\prime(z)-1/z^2$, with $\Psi^{\prime}(1/2)= \frac{\pi^2}{2},  \Psi^{\prime}(1)= \frac{\pi^2}{6}$.  When $s$ or $\lambda$ increases the number of bound states possessed by these two potentials increase and the ground state lies deeper and deeper. Interestingly, in the limit when $s,\lambda \rightarrow \infty$, both the uncertainty products (3,4) can be readily checked to tend to the minimal value of $\hbar/2$ for ground states. Perhaps, this could be a common feature of one-dimensional potentials possessing finite number of discrete eigenvalues. 

Let $\phi(p)$ the momentum space representation of the eigenfunction which is the Fourier transform  of the eigenstate $\psi(x)$, then  physical quantities like $<x>, \Delta x, <p> , \Delta p$ and $U$ are known [1-3,6-10] to be independent of the representation (use  $\phi(p)$ or $\psi(x)$). The proof of this is often  left as an exercise.  In principle one can do all calculations in momentum space just as well (though not always as easily) as in position space.

Now if we are given  $\psi(x)$ and $\phi(x)$ (notice that it is $x$ and not $p$ which is the argument of $\phi$), here, we point out that  the well known equivalence of results using $\psi(x)$ or its momentum representation, $\phi(p)$, manifests in  
\begin{equation}
(\Delta x)_{\psi(x)}= (\Delta p)_{\phi(x)}, \quad
(\Delta p)_{\psi(x)}= (\Delta x)_{\phi(x)}, \Rightarrow
U_{\psi(x)}=U_{\phi(x)}.
\end{equation}
Therefore, if one finds $U_{\psi(x)}$, one has found $U_{\phi(x)}$ as well. However, it may turn out that one may not be doable as easily as the other one. In Table I, we display several pairs of ground states $\psi_0(x)$ and $\phi_0(x)$ which may look similar (see
row nos. 2 and 3) or dissimilar (see row nos. 1, 4-7)
but they essentially give rise to the same value for $U$. For the proof of the equivalence in (5) see the Appendix.
\begin{table}

	\centering
	
		\begin{tabular}{|c||c||c||c||c||c|}
		\hline
		S.N. & ~~~~$V(x)$~~~ & ~~~~$\psi_0(x)$~~~~ & ~~~~~$\phi_0(x)$~~~~~~ & ~~~~$U$~~~~~&~~Ref.~~  \\
		\hline
		\hline
		1 & IDW & $ \sqrt{\frac{2}{a}} \sin \pi x/a, 0 \le x\le a;$ & $2\sqrt{a\pi} \frac{1+e^{iax}}{(\pi^2-a^2x^2)}$ & $\frac{\hbar}{2}\sqrt{\frac{\pi^2-6}{3}}$ & [3] \\
		& & 0, \quad otherwise &  & & \\
		\hline
		2 & H.O& $\frac{e^{-x^2/2}}{\pi^{1/4}}$& $\frac{e^{-x^2/2}}{\pi^{1/4}}$& $\frac{\hbar}{2}$ & [1-3,6-10] \\
		\hline
		3 & $V_{SRM}(s= 1)$ & $\frac{1} {\sqrt{2}} \mbox{sech} x$ & $\frac{\sqrt{\pi}}{2} \mbox{sech}\frac{x\pi}{2}$ & $\frac{\hbar \pi}{6}$ & Eq.(3) [4]\\
 \hline 
 4 & $V_{SRM} (s=2)$& $\frac{\sqrt{3}}{2} \mbox{sech}^2 x$ & $\sqrt{\frac{3\pi}{8}} x~ \mbox{cosech} \frac{x \pi}{2}$ & $\hbar \sqrt{\frac{\pi^2-6}{15}}$ & Eq.(3) [4]\\
 \hline
 5 & $ V_{M} (\lambda=1)$ & $\sqrt{2}~ e^{-(e^x-x/2)}$ &$ \frac{1}{\sqrt{\pi}} \Gamma(\frac{1}{2}+ix)$& $\frac{\hbar \pi}{2\sqrt{6}}$ & Eq.(4) [4] \\
 \hline
 6 & $-V_0 \delta(x)$ & $\frac{1}{\sqrt{a}}e^{-|x|/a}$ & $\sqrt{\frac{2a}{\pi}} \frac {1}{1+a^2x^2}$ &  $\frac{\hbar}{\sqrt{2}}$ & Eq.(12)\\
 \hline
 7 & Eq.(19) & $\sqrt{\frac{3}{2a}} (1-|x|/a), |x|\le a$
& $\sqrt{\frac{3a}{\pi}}\left(\frac{\sin ax/2} {ax/2}\right)^2$ & $\sqrt{\frac{3}{10}} \hbar$ & Eq.(25) \\
& & 0, \quad $|x|>a$ & & & \\
 \hline
      
		\end{tabular}
		\caption{The ground states, $\psi_0(x)$, of various potentials, $V(x)$, the corresponding  $\phi_0(x)$ and the position momentum uncertainty products,$U$. $\phi_0(p)$ is the Fourier transform of $\psi_0(x)$. Here IDW is infinitely deep well potential 
of width $a$, H.O is Harmonic oscillator, $V(x)= x^2/4$, $V_{SRM}(x)$ and $V_M(x)$ are given above the Eq. (3). 
The pairs of wave functions $\psi_0(x)$ and $\phi_0(x)$ which are Fourier transform of each other may look incidentally similar (see row 2 and 3) or generally dissimilar (see row 1,4-7) they would however give rise to the same value for $U$.}
\end{table}
Next, in the following we present the determination of the uncertainty product for two special eigenstates: ${\bf Case~(I)}$ and ${\bf Case ~(II)}$.\\ \\
{\bf Case (I)- Dirac delta well:} This potential $V(x)=-V_0 \delta(x), V_0>0$
is well known to have a single bound state at $E=-\frac{m V_0^2}{2\hbar^2}$ and
 its normalized eigenfunction is given as [6-9]
\begin{equation}
\psi_0(x)=\sqrt{\alpha} e^{-\alpha |x|}, \quad \alpha=\frac{mV_0}{\hbar^2}. 
\end{equation}
The expectation value of $x$ for this state vanishes as it is an
even parity state. The expectation value of $x^2$ for this state is given
\begin{equation}
<\psi_0|x^2|\psi_0>= \alpha \int_{-\infty}^{\infty} x^2 e^{-2 \alpha|x|}~ dx=\frac{1}{2\alpha^2}.
\end{equation}
When the momentum operator $p=-i\hbar \frac{d}{dx}$ operates over $\psi_0$, we have
\begin{equation}
p\psi_0(x) = i\hbar \alpha \sqrt{\alpha} e^{-\alpha |x|} \frac{d|x|}{dx}=  i \hbar \alpha \sqrt{\alpha} e^{-\alpha |x|}~ \mbox{sgn}(x), 
\end{equation}
where $\mbox{sgn}(x)$ is called signum function which is defined as
\begin{equation} 
\mbox{sgn}(x)=\left\{\begin{array}{lcr}
-1, & &  x < 0,\\
0, & & x=0,\\
+1, & &  x> 0.
\end{array}
\right.
\end{equation}
So it follows that
\begin{equation}
<\psi_0|p|\psi_0>=<\psi_0(x)|p\psi_0(x)>=i\alpha^2 \hbar \int_{-\infty}^{\infty} e^{-2\alpha|x|}~ \mbox{sgn}(x)~dx
\end{equation}
vanishes as $\mbox{sgn}(x)$ is an odd function. This conforms to the fact that for a bound state, the expectation value of momentum is zero. Next we find 
\begin{equation}
<p\psi_0|p\psi_0>=\alpha^3 \hbar^2 \int_{-\infty}^{\infty} e^{-2\alpha |x|} (\mbox{sgn}(x))^2 ~dx= \alpha^2 \hbar^2,
\end{equation}
as $(\mbox{sgn}(x))^2=1$, except at $x=0$. The momentum being
a Hermitian operator, Eq. (11) gives nothing but $<\psi_0|p^2|\psi_0>$, giving us from Eqs. (2,7,11)
\begin{equation}
U_{\psi_0}=\Delta x \Delta p =\frac{\hbar}{\sqrt{2}}. 
\end{equation}
We can also write
\begin{equation}
\mbox{sgn}(x)=2\theta(x)-1,
\end{equation}
where $\theta(x)$ is called the Heaviside step function which is defined as [1,2,6,7]
\begin{equation} 
\theta(x)=\left\{ \begin{array}{lcr}
1, & &  x\ge 0\\
0, & & x<0
\end{array}
\right.
\end{equation}
and the Dirac Delta function, $\delta(x)$, is defined as [1,2,6,7]
\begin{equation}
\delta(x)=\frac{d \theta(x)}{dx}.
\end{equation}
Differentiating  Eq.(8) with respect to $x$ and multiplying it by $-i\hbar$, we can write 
\begin{equation}
p^2 \psi_0 = -\hbar^2 \sqrt{\alpha} [\alpha^2 e^{-\alpha |x|} (\mbox{sgn}(x))^2-2 \alpha \delta(x) e^{-\alpha |x|}].
\end{equation}
We can get $<\psi_0|p^2|\psi_0>$ alternatively as
\begin{equation}
<\psi_0|p^2|\psi_0>=-\hbar^2 \alpha \left [ \alpha^2\int_{-\infty}^{\infty}  e^{-2 \alpha |x|}~dx -2\alpha \int_{-\infty}^{\infty} e^{-2\alpha|x|} \delta(x)~dx \right] =
\hbar^2 \alpha^2.
\end{equation}
The second integral in (17) is 1 using the property that [1,2,6,7]
\begin{equation}
\int_{-\infty}^{\infty} f(x) \delta(x) dx= f(0)= \int_{-a}^{b} f(x) \delta(x) dx, \quad ab>0.
\end{equation}
We would like to mention that the Ex-4 on page 47 in Ref. [8] and Ex 1.1 on page 16 in Ref. [9]  finding the uncertainty product for a hypothetical Lorentzian eigenstate $\psi_0(x)=A[x^2+\alpha^2]^{-1}$ leads to the same result as (12). This is only a verification of the equivalence (5). It may be verified (see the row no. 6 in the Table 1) that for the Lorentzian state $\psi_0(x)$, $\phi_0(x)$ is nothing but the symmetric exponential function appearing in Eq. (6).\\ \\
{\bf Case(ii)-
Dirac delta between two rigid walls:} This potential
\begin{equation} 
V(x)=\left\{ \begin{array}{lcr}
\infty, & &  |x|\ge a\\
-V_0 \delta(x), & & |x|<a
\end{array}
\right. \quad V_0>0
\end{equation}
possesses [11,12] an interesting zero-energy, zero-curvature bound state conditionally when $\frac{maV_0}  {\hbar^2}=1$. The normalized ground state  can be written as as 
\begin{equation}
\psi_0(x)= \sqrt{\frac{3}{2a}}\left( 1-\frac{|x|}{a} \right), \quad -a \le x\le a.
\end{equation}
Notice that $\psi_0(x)$ vanishes at $x=\pm a$ due to the presence of rigid walls. Due to the symmetry of this state $<x>=0$ and $<x^2>$ can be found as
\begin{equation}
<\psi_0|x^2|\psi_0>=\frac{3}{2 a} \int_{-a}^{a} x^2 \left(1-\frac{|x|}{a}\right)^2~dx=\frac{a^2}{10}.
\end{equation}
The action of $p$ over $\psi_0(x)$ is
\begin{equation}
p \psi_0(x)= i\hbar \sqrt{\frac{3}{2a^3}}~ \mbox{sgn}(x),
\end{equation}
which being an odd function gives the expectation value of $p$ as
\begin{equation}
<\psi_0|p\psi_0>=i\hbar \frac{3}{2a^3}~\int_{-a}^{a} \left(1-\frac{|x|}{a}\right ) \mbox{sgn}(x)~dx=0.
\end{equation}
However, we have 
\begin{equation}
<p\psi_0|p\psi_0>= \hbar^2\frac {3}{2a^3} \int_{-a}^{a} (\mbox{sgn}(x))^2 ~dx= \frac{3 \hbar^2}{a^2},
\end{equation}
which is nothing but $<\psi_0|p^2\psi_0>$, then from Eqs. (2,21,24) we have the uncertainty product for the zero-energy
and zero curvature eigenstate [11,12] as
\begin{equation}
U_{\psi_0}=\Delta x \Delta p= \sqrt{\frac{3}{10}}\hbar,
\end{equation}
which is approximately $0.5477 \hbar$ greater than $\hbar/2$ 
and rightly so.
Alternatively, if we differentiate Eq. (22) with respect to $x$ and multiply it by $-i\hbar$, then we can write 
\begin{equation}
p^2\psi_0= 2\hbar^2 \sqrt{\frac{3}{2a^3}}~ \delta(x). 
\end{equation}
We recover the result (24) as
\begin{equation}
<\psi_0|p^2\psi_0>=2\hbar^2 \frac{3}{2a^2} \int_{-a}^{a} \left(
1-\frac{|x|}{a}\right)\delta(x)~ dx= \frac{3 \hbar^2} {a^2},
\end{equation}
by using (18).
In Table I, see that the Fourier-transform of the eigenstate, $\psi_0(x)$, given by Eq. (20) is $\phi_0(p)=\sqrt{\frac{3a}{\pi}} \left (\frac{\sin ap/2}  {ap/2}\right)^2, -\infty< p <\infty$. By using $\phi_0(p)$ the result (25) can be recovered again but by carrying out apparently different integrations which may not be easier to do.

Lastly, we would like to remark that these two special eigenstates of two potentials could be a new addition to the exercises of finding the uncertainty products and confirming that in one dimension these are greater than $\hbar/2$. Our exposition that the ground state  attains the minimum uncertainty product ($\frac{\hbar} {2}$) when the depth of the potential tends to infinity requires further confirmation.  
Students may find  the equivalence of the uncertainty product revealed here, interesting and enriching. The handling of the notional functions such as $\mbox{sgn}(x), \theta(x)$ and $\delta(x)$ here is also instructive.

\renewcommand{\theequation}{A-\arabic{equation}}
\setcounter{equation}{0}
\section*{Appendix}
The basic definition of the Dirac delta function is
\begin{equation}
\frac{1}{2\pi} \int_{-\infty}^{\infty} e^{i(a_1-a_2)b} ~db= \delta(a_1-a_2).
\end{equation}
We can differentiate (A-1) with respect to $a_1$, $n$ times, to get the definition of derivatives of the Dirac delta function as
\begin{equation}
\frac{(i)^n}{2 \pi} \int_{-\infty}^{\infty} ~b^n~ e^{i(a_1-a_2)b} ~db= \left(\frac{\partial}{\partial a_1}\right)^n \delta (a_1-a_2)=\delta^{(n)}(a_1-a_2)
\end{equation}
Using integration by parts, we can write
\begin{equation}
\int_{-\infty}^{\infty} f(a_1) \delta^{(1)} (a_1-a_2) da_1=  f(a_1) \delta (a_1-a_2)|_{-\infty}^{\infty}-\int _{-\infty}^{\infty}f^{(1)}(a_1) \delta(a_1-a_2) da_1= -f^{(1)}(a_2).
\end{equation}
Similarly repeated integrations by parts lead to  
\begin{equation}
\int_{-\infty}^{\infty} f(a_1) \delta^{(n)} (a_1-a_2) da_1=-\int_{-\infty}^{\infty}f^{(n)}(a_1) \delta(a_1-a_2) da_1= (-1)^nf^{(n)}(a_2).
\end{equation}
Now let $\psi(x)$ be an eigenstate whose Fourier transform or momentum representation is $\phi(p)$.
So we can write
\begin{equation}
\phi(p)=\frac{1}{\sqrt{2\pi\hbar}} \int_{-\infty}^{\infty}
e^{-ipx/\hbar} \psi(x)~ dx, \quad \mbox{or}, \quad \psi(x)=\frac{1}{\sqrt{2 \pi\hbar}} \int_{-\infty}^{\infty}
e^{ipx/\hbar} \phi(p)~ dp. 
\end{equation}
We find $<\psi(x)|x|\psi(x)>$ denoting it as
\begin{equation}
<x>_{\psi(x)}=\int_{-\infty}^{\infty} \psi^*(x) ~x~\psi(x)~dx=\int_{-\infty}^{\infty}\int_{-\infty}^{\infty}\int_{-\infty}^{\infty} \phi^*(p_1) x e^{-i(p_1-p_2)x/\hbar} \phi(p_2)~dx~dp_1~dp_2
\end{equation} 
Carrying out the $x-$ integration  using (A-2) and the carrying out $p_1-$integration using (A-3), we get
$<\psi|x|\psi>$ which we denote 
\begin{equation}
<x>_{\psi(x)}=\int_{-\infty}^{\infty} \phi^*(p_2) ~(-i\hbar) \frac{\partial}{\partial p_2}~\phi(p_2)~dp_2=
\int_{-\infty}^{\infty} \phi^*(x) ~(-i\hbar) \frac{\partial}{\partial x}~\phi(x)~dx=<p>_{\phi(x)}.
\end{equation}
Normally, one would like to term the second part in the above equations as $<\phi(p)|x|\phi(p)>$, namely the expectation value of $x$ in momentum space, which is the same as $<\psi(x)|x|\psi(x)>$. It is here we depart from this
and instruct that the third part in the above equations is merely due to the fact that in a definite integral the name of the variable is only dummy so $p_2$ could be changed to $x$. Then follows the last part wherein we identify $-i\hbar\frac{\partial}{\partial x}$ as momentum operator $p$. Similarly, we can prove that $<x^2>_{\psi(x)}=<p^2>_{\phi(x)}, <p>_{\psi(x)}=<x>_{\phi(x)}, <x^2>_{\phi(x)}=<p^2>_{\psi(x)}.$
Hence the claim in (5) is proved. \\ \\
{\bf Acknowledgment:} We thank Dr. V. M. Datar for his support and interest in this work. IY would like to thank the Physics batch-mates of 57$^{th}$ batch of Training School of BARC for fond memories.

\end{document}